\documentclass[]{aa}
\usepackage{graphicx}
\usepackage{multirow}
\usepackage{subfig}
\usepackage{svg}
\usepackage{ulem}
\usepackage[varg]{txfonts}
\newcommand{\beq}{\begin{equation}}
\newcommand{\eeq}{\end{equation}}
\newcommand{\ben}{\begin{eqnarray}}
\newcommand{\een}{\end{eqnarray}}
\newcommand{\Fig}[1]{Fig.~\ref{#1}}
\newcommand{\fig}[1]{\Fig{#1}}
\graphicspath{{figures/}}
\begin{document}

\title{The effect of dynamic temperatures on pebble dynamics and planet formation} 
\author{Areli Castrejon\inst{\ref{inst1},\ref{inst2},\ref{inst3}}
\and Michiel Min\inst{\ref{inst2}}
\and Inga Kamp\inst{\ref{inst1}}
\and Uffe Gråe Jørgensen\inst{\ref{inst3}}}
\institute{Kapteyn Astronomical Institute, University of Groningen, Groningen, The Netherlands\label{inst1}
\and Netherlands Space Research Institute (SRON), Leiden, The Netherlands\label{inst2}
\and Centre for ExoLife Sciences, Niels Bohr Institute, University of Copenhagen, Copenhagen, Denmark\label{inst3}}
\abstract{To date, more than 5000 exoplanets have been discovered. The large majority of these planets have a mass between 1 and 17 $M_\oplus$, making them so-called super-Earths and mini-Neptunes. The exact formation process for this abundant planet population has not yet been fully constrained.}
{Recent studies on the formation of these planets make various assumptions with regard to the disk. The primary mass budget, held in pebbles, is either assumed to have a constant size or is parametrized as a flux. Simplifications of the temperature structure, in the form of a static power law, do not consider the temperature evolution and high magnitudes of heating in the inner part of the disk. In this study, we aim to investigate the effect these simplifications of temperature and pebble sizes have on the pebble densities and resulting planet populations.}
{To constrain the timescales needed to form super-Earths, we developed a model for exploring a large parameter space. We included the effect of two different temperature prescriptions on a viscously accreting and spreading disk. We formed a pebble reservoir utilizing a simplified conversion timescale with a time- and radially dependent Stokes number for the dust. We then tracked the temporal evolution of the surface densities of gas, dust, and pebbles. Pebbles were allowed to drift and be accreted onto a growing protoplanet. As a planet grows, it exerts a torque on the disk, carving out a gap and affecting the pebble drift, before halting the growth of the planet.}
{We find that viscous heating has a significant effect on the resulting mass populations, with the static power law showing smaller planets within 10 AU. Inside the dust-sublimation line, usually within 0.5 AU, planet formation is reduced due to the loss of planet-forming material. Our model replicates observed planet masses between Earth and mini-Neptune sizes at all radial locations, with the most massive planets growing in the intermediate turbulence of $\alpha=10^{-3}$.}
{We conclude that a self-consistent treatment of temperature, with the inclusion of a dust-sublimation line, is important and could explain the high occurrence of super-Earths at short orbital separations.}

\maketitle
%
%

%
%
%
%
\section{Introduction}
To date, more than 5000 exoplanets have been discovered \citep{Akeson+13,Petigura+13,Mulders+18}. The large majority of these planets are in the intermediate-mass range, between 1 and 17 $M_\oplus$, and are thus so-called super-Earths and mini-Neptunes \citep{Schneider+11,Bean+21}. Despite the abundance of these planets in our galactic neighborhood, their formation processes have not yet been fully constrained. \par 
One theory for their formation is the core accretion scenario, whereby larger planetesimals serve as planetary embryos and accrete other planetesimals, thereby forming the core of giant planets \citep{Pollack+96}. However, the timescale for planetary growth via planetesimal accretion is often longer than the lifetime of a protoplanetary disk \citep{Pollack+96,LambrechtsJohansen12,JohansenLambrechts17,Ormel+17}.
One solution to this timescale problem is accretion from smaller materials that are readily available in the disk. These smaller materials can be efficiently accreted once a planetesimal is formed, through processes such as the streaming instability \citep{Johansen+15,LiYoudin21}. The formation of larger particles begins with collisions of interstellar size dust, resulting in coagulation \citep{Brauer+08,Zsom+10,Birnstiel+12}. However, particles do not just keep growing incessantly; there are various disturbances that play a role in the formation of planetesimals. Planetesimal formation is tempered by fragmentation, bouncing, and radial-drift barriers \citep{Brauer+08,Birnstiel+12}. This hindrance in the formation of planetesimals results in an abundance of millimeter- to centimeter-sized particles. These larger particles, called pebbles, become an important part of the planet formation process.
Reservoirs of material in the outer disk constantly replace the material being lost to planetary accretion and accretion onto the star \citep{Weidenschilling77}. The shorter timescale of accretion for pebbles allows planets to grow quickly, before the dissipation of the dusty and gaseous disk \citep{LambrechtsJohansen14, Bitsch+18, Lambrechts+19}. As pebbles grow, they experience a headwind drag from the gaseous disk, which causes them to lose angular momentum and drift inward. The constant flux of pebbles paired with a large embryo-accretion cross section allows for efficient accretion onto a protoplanet \citep{Whipple72, Ormel17}. The growth from dust grains to pebbles and the subsequent planet formation processes are discussed in depth in the reviews by \cite{LiuJi20} and \cite{Drazkowska+23}.\par
Various studies on pebble accretion show that super-Earths form easily within the lifetimes of protoplanetary disks \citep{Bitsch+19b, Liu+19, Venturini+20, SavvidouBitsch23}. Recent works on super-Earth formation show that the resulting mass for super-Earth planets is dependent on the pebble flux through the inner disk \citep{Bitsch+19a, Liu+19}. Other works find that the pebble isolation mass, which is highly dependent on the temperature structure, has a significant effect on the final mass of the planet formed \citep{Bitsch19,SavvidouBitsch21}. Recently, \citet{Venturini+20} found that the largest planets formed within the water ice line are at most 5 Earth masses.\par
In this work, we aim to investigate the effects that viscous heating with dust sublimation, self-consistent pebble sizes, and 1D gap formation have on the formation of super-Earth populations. Our main motivation for this study was to better understand pebble drift, which is set by the headwind velocity, a quantity that is dependent on the disk temperature. A power law for the temperature is commonly used, representing a disk heated only by the central star \citep{Ida+16}. However, temperatures in the inner 5 AU of the disk can reach higher levels due to viscous heating \citep{Min+11, Bitsch+15, Ida+16}. At high temperatures, dust can sublimate, depending on its composition. As the dust sublimates \citep{Kobayashi+11}, it removes opacity from the disk, causing the temperature to drop and remain constant, independent of radius, leading to a so-called dust thermostat \citep{Min+11}. These changes in temperature structure should likely affect the radial evolution of pebble densities.\par
In this work, we focused on the growth of planets via pebble accretion onto a preformed protoplanetary embryo. We followed the growth of the planet and its corresponding gap carving in the gaseous disk \citep{LinPapaloizou86, Alibert+05}. Earlier 1D studies did not investigate how a gradual gap-formation process affects the pebble flux and thus planet buildup. Here, we investigated how this process compares to works that used the pebble isolation mass to halt planet growth. We followed the temperature evolution of the disk using two different temperature prescriptions: (i) a model that includes a viscous and passively heated disk with a dust thermostat and dust removal and (ii) a passively heated power-law model. We investigated the effect that different temperature prescriptions have on pebble evolution and subsequently on the formation of planets. Based on our findings, we performed a parameter study, varying the viscosity, embryo location, and insertion time.\par
This paper is structured as follows. In Sect. \ref{sect:method}, we explain our semi-analytic model in detail. This includes the different temperature prescriptions, the formation of pebbles from dust while including growth limits, how a planet carves a gap in the disk, and the initial timeline for planet formation. In Sect. \ref{sect:results}, we present the results from our temperature prescriptions on pebble growth and transfer. We then interpret the implications for planet formation and pebble formation in Sect. \ref{sect:discussion}. Finally, we summarize our findings in Sect. \ref{sect:conclusions}.
%
%
\section{Methods}
\label{sect:method}
Exploring the parameter space for planet formation requires a swift model with moderate complexity. As such, we developed a computationally fast model for the temporal evolution of gas, dust, and pebble densities, while intending to capture most aspects of the evolution in a realistic manner. We used a simplified recipe to describe how dust is turned into pebbles as a function of time and radius. We inserted Moon-mass protoplanets into the disk and let them grow through pebble accretion. Once a planet grows large enough, it can carve a gap in the disk, leading to a pressure bump and the slowdown of pebble accretion. We ran a suite of models, exploring two different assumptions for disk temperatures. Within each temperature prescription, the variable quantities were the disk viscosity and the initial time of embryo insertion.
\subsection{Protoplanetary disk model}
We began with the equations of conservation of mass and angular momentum for a gaseous disk \citep{Lynden-BellPringle74},  
\ben
\frac{\partial\Sigma_{\rm g}}{\partial t} = \frac{1}{r} \frac{\partial}{\partial r} \left[ \Sigma_{\rm g} r \left( u_{\rm g} + u_{\tau}\right) \right],
\een
where the gas velocity, $u_{\rm g}$, is given by
\beq
u_{\rm g} = \frac{-3}{\Sigma_{\rm g}\sqrt{r}} \frac{\partial}{\partial r} (\Sigma_{\rm g} \nu r),
\eeq
and $u_{\tau}$ is a velocity term arising from torques once a planet is large enough to disturb the disk. This term, $u_{\tau}$,  is further discussed in Sect. \ref{sect:planet_disk_interactions}.
Here, $\Sigma_{\rm{g}}$ is the gaseous surface density and $r$ is the radial location in the disk.
The viscosity in the disk, $\nu,$ is given by
\beq
\nu = \alpha c_{\rm s} H,
\eeq
where $\alpha$ is a dimensionless parameter that describes the turbulent strength \citep{ShakuraSunyaev73}. The sound speed, $c_{\rm s}$, and scale height, $H$, are given by
\beq
c_{\rm{s}} = \sqrt{\frac{k_{\rm{B}}T}{\mu m_{\rm p}}}\\
\eeq
and
\beq
H = \frac{c_{\rm{s}}}{\Omega}, 
\eeq
where $k_{\rm B}$ is the Boltzmann constant, $T$ is the midplane temperature in the disk, $\mu$ is the mean molecular weight of the gas (assumed to be molecular), and $m_{\rm p}$ is the proton mass.
The Keplerian frequency, $\Omega,$ is given by
\beq
\Omega =  \sqrt{\frac{GM_*}{r^3}},
\eeq
with $G$ and $M_*$ as the gravitational constant and mass of the star, respectively.
We set our initial gas surface density as a function of radial distance, $r$, to a tapering power law, namely the one used in \citet{Lynden-BellPringle74},
\beq
\Sigma_{\rm{g,0}} = (2 - \gamma) \frac{M_{\rm{disk}}}{2 \pi {r_{\rm{c}}}^2} \left(\frac{r}{r_{\rm{c}}}\right)^{-\gamma}\exp{\left(-\frac{r}{r_{\rm{c}}}\right)},
\eeq
where $M_{\rm{disk}}$ is the disk mass, $r_{\rm{c}}$ is the characteristic radius, and $\gamma$ is the power-law exponent. The initial dust surface density was set to 
\beq
\Sigma_{\rm{d,0}} = Z_0 \cdot \Sigma_{\rm{g,0}},
\eeq
where $Z_0$ is the initial dust-to-gas ratio.
\subsection{Temperature structure}
We set out to investigate the effect that different temperature structures have on the evolution of a protoplanetary disk and, by association, on the planet formation process. Previous studies on planet formation have treated the temperature structure as being dominated by only stellar irradiation \citep{Armitage10}. To investigate the effect of this assumption, we investigated two different temperature structures.
We considered heating from stellar irradiation and viscous stresses, following the analytical temperature model in \citet{Min+11}. The temperature is dominated by viscous heating  in the inner disk and by stellar irradiation in the outer disk. Upon determining the initial gas surface density, we calculated the temperature in the midplane of the disk. The equation for the viscous temperature is
\beq
T_{\rm{vis}} = \left[\frac{27}{128}\left(\frac{\Sigma_{\rm{g}} \kappa_{\rm{R}} \alpha k_{\rm{B}} Z \Omega}{\sigma \mu m_{\rm{p}}}\right)\right]^{1/3},
\label{temp_viscous}
\eeq
where $\kappa_{\rm{R}}$ is the Rosseland mean opacity, and $Z$ is the local dust-to-gas ratio. This dust-to-gas ratio changes over time and differs from our initial dust-to-gas ratio, $Z_0$.
The temperature from irradiation is 
\beq
T_{\rm{irr}} = \left[\varphi\frac{L_*}{4\pi\sigma r^2}\right]^{1/4},
\label{temp_irr}
\eeq
where $\varphi$ is the grazing angle, and $L_*$ is the stellar luminosity. 
We added the contributions from the viscous and irradiative energies to find our final midplane temperature,
\beq
T^4 =  T_{\rm{irr}}^4 + T_{\rm{vis}}^4.
\eeq
\subsubsection{Standard model}
As reported by \citet{Min+11}, the temperature in the inner disk reaches high enough temperatures to exceed the silicate sublimation temperature. The balance between the deposition and sublimation of matter causes a thermostat effect, limiting the midplane temperature to the silicate sublimation threshold. When temperatures reach a value higher than the silicate sublimation temperature given by Eq. \eqref{temp_viscous}, dust sublimation will lower the temperature until a balance is reached at $\approx$ 1500 K \citep{Kobayashi+11}. This sublimation process reduces the available material for planet formation and results in a flat temperature structure in the inner parts of the disk. To reproduce this flat temperature in the inner disk, we calculated the $Z$ needed for $T$=1500 K and then removed the dust and added it to our gas component. We refer to this reduced dust and temperature model as our standard model.

\subsubsection{Power-law model}
To be able to compare our results to previous works, we also used a static power-law temperature of
\beq
T_{\rm{const}} = 186 \ \left(\frac{r}{\rm{AU}} \right)^{-1/2} \ \rm{K}.
\eeq
We chose this temperature structure to coincide with the temperature profile in the outer disk regions of our previous models, as seen in Fig. \ref{fig:temp_comp_powerlaw}. This equation is also a solution for Eq. \eqref{temp_irr} using our model parameters. We assume that this temperature structure does not evolve in time and refer to this simplified temperature prescription as our power-law model.
\subsection{Growing from dust to pebbles}
In a disk, not all dust particles are identical. Particles of different sizes behave differently based on their Stokes number, a quantity that defines the particles' coupling to the gas (see, e.g., \citealt{Ormel17}). The Stokes number in the Epstein drag regime \citep{Brauer+08} is given by
\beq
{\rm St} = \frac{\pi a \rho_{\rm{s}}}{2 \Sigma_{\rm{g}}},
\eeq
where $\rho_{\rm{s}}$ is the material density of the grain, and $a$ is the size of the grain. \citet{Ormel17} defines pebbles as falling within the range $10^{-2}\leq\rm{St}<1$. Therefore, pebbles can range in physical size from millimeter- to centimeter-sized particles \citep{Perez+15}.\par
Pebbles are a vital component in planet formation since they experience a gas drag that slows down their motions, allowing for faster protoplanetary embryo growth \citep{Ormel17}. We followed previous works that used a simplified approach to calculate the maximum sizes of grains \citep{Birnstiel+12,DrazkowskaAlibert17,Venturini+20}. This simplified approach determines the available pebble population as a function of time and radius.
We subdivided our disk solids into two distinct populations, small grains coupled to the gaseous disk motions (dust) and larger grains that can drift through the disk (pebbles).
We began with the equation to grow to a determined size from \citet{Brauer+08},
\beq
a = a_0 \exp \left(\frac{\Omega\Sigma_{\rm d} t}{\Sigma_{\rm g}}\right).
\label{eqn:growth_timescale}
\eeq
The equation states that the growth of a particle of size, $a$, is dependent on the initial particle size, $a_0$, and the exponent of the local dust-to-gas ratio, the Keplerian frequency, and the time elapsed. This assumption is valid for particles with a Stokes number less than unity.
\par
However, the growth of grains does not continue in this way up to planetary sizes. Particle growth is subject to limiting factors, two of which are determined by turbulent fragmentation and radial drift. In the fragmentation case, dust grains can fragment into smaller particles due to the relative motions caused by turbulence in the disk. In this case, the maximum grain size is given by
\beq
a_{\rm frag} = \frac{2  \Sigma_{\rm g}  v_{\rm frag}^2}{3 \pi \rho_{\rm s} \alpha c_{\rm s}^2},
\eeq
where $v_{\rm frag}$ is a free parameter than can range from 100 to $10^3 \ [\rm cm/s]$ \citep{GundlachBlum15,MusiolikWurm19}. The second limiting factor is the balance between the growth timescale and the drift timescale. This arises because the timescale for particle drift can be similar to or less than the time it takes to grow to a larger grain size. This particle size is given by
\beq
a_{\rm drift} = \frac{2 \Sigma_{\rm d} v_{\rm k}^2}{\pi  \rho_{\rm s} c_{\rm s}^2} \left|\frac{\partial \log P}{\partial \log r}\right|^{-1}.
\eeq
Based on these three criteria, we chose the critical size of our pebbles to be
\beq
    a_{\rm crit}= {\rm min}\left\{a_{\rm drift}, a_{\rm frag}, a_0 \exp \left(\frac{\Omega\Sigma_{\rm d}t}{\Sigma_{\rm g}}\right) \right\}.
\eeq
Knowing the critical size, we arrived at an analytical solution for the critical time to transform dust into pebbles of a maximum size using Eq. \eqref{eqn:growth_timescale}. Plugging our critical size, $a_{\rm crit}$, into $a$, we were able to solve for $t_{\rm crit}$:
\beq
t_{\rm{crit}} = {\log{\left(\frac{a_{\rm crit}}{a_{\rm 0}}\right)}}\frac{\Sigma_{\rm g}}{\Omega\Sigma_{\rm d}}.
\eeq
Over time, we populated our pebble surface density through a source term dependent on the dust surface density,
\beq
S_{\rm p} = \Sigma_{\rm d} \cdot \left[ 1 - {\rm exp} \left( \frac{-t}{t_{\rm crit}} \right) \right].
\eeq
This allowed for the growth transition from dust to pebbles to proceed smoothly. Theoretically, as the model progresses, the exponential reaches a value of unity, resulting in a total conversion of dust into pebbles.\par
In a protoplanetary disk, however, the efficiency of dust-to-pebble conversion will not be fully efficient since fragmentational collisions will erode pebbles, replenishing the small grains. We assumed that the fragmentation of pebbles results in grains that are 1 micron in size. Hence, we did not fully deplete our dust population and opted to keep a certain dust fraction. This simplification allows the viscous temperature in the disk to remain high, as the replenishment of small grains keeps the opacity high. We chose to limit our particle populations, keeping 75\% as pebbles and 25\% as dust, which we achieved by first calculating the ratio of pebbles making up the total solid mass using
\beq
\Sigma_{\rm tot}=\frac{\Sigma_{\rm p}}{\Sigma_{\rm p} + \Sigma_{\rm d}}.
\eeq
If this quantity was larger than 75\%, we removed the excess pebbles and added them back into the dust surface density. This ensured enough dust to simulate viscous heating and an evolving temperature structure in the inner disk.
\subsection{Pebble evolution}
Pebbles orbiting a gaseous disk attempt to move on Keplerian orbits but experience an aerodynamic headwind due to gas moving at sub-Keplerian speed. Pebbles drift toward the inner disk according to the equation from \cite{Birnstiel+10,Birnstiel+12},
\ben
\frac{\partial \Sigma_{\rm{p}}}{\partial t} = -\frac{1}{r} \frac{\partial}{\partial r} \left[ r \left( \Sigma_{\rm p} u_{\rm r} - D_{\rm d}\Sigma_{\rm{p}} \frac{\partial}{\partial r} \left(  \frac{\Sigma_{\rm{p}}}{\Sigma_{\rm{g}}}\right) \right) \right] + S_{\rm p},
\een
where $\Sigma_{\rm{p}}$ is the pebble surface density, and $D_{\rm d}$ is the dust diffusion coefficient, usually given by this equation from \cite{YoudinLithwick07}:
\beq
D_{\rm d} = \frac{\nu}{1+{\rm St}^2}. 
\eeq
\par
The radial velocity of the pebbles, $u_{\rm r}$, contains two contributions. The first term is the drag term due to the gas velocity, $u_{\rm g}$, which drags the pebbles to an extent, as they do not fully decouple from the gas. When a planet was introduced, we included the velocity due to the planetary torque, $u_\tau$. The second contribution is the radial drift velocity, $u_{\rm n}$, with respect to the gas. Gas in a protoplanetary disk experiences the force of its own pressure support, which causes it to move in a sub-Keplerian fashion. This leads to moving pebbles experiencing a constant headwind, which causes them to lose angular momentum and drift inward. Both of these velocities scale depending on the Stokes number of the particles, such that the total radial velocity is given by
\beq
u_{\rm r} = \frac{1}{1+{\rm St}^2} (u_{\rm g} + u_{\tau}) - \frac{{\rm St}}{1+{\rm St}^2} 2 u_{\rm n},
\eeq
where the drift velocity, $u_{\rm n}$, is given by
\beq
u_{\rm{n}} = -\frac{c_{\rm s}^2}{2 \Omega r}\frac{\partial\log P}{\partial\log r}.
\eeq
The midplane pressure, $P$, that determines the drift velocity is given in our model by
\beq
P  = \rho_{\rm{g}} c_{\rm{s}}^2,
\eeq
where the midplane gas density, $\rho_{\rm{g}}$, is
\beq
\rho_{\rm{g}} = \frac{\Sigma_{\rm{g}}}{\sqrt{2 \pi}H}.
\eeq
\subsection{Planet formation}
To model the planet formation process, we inserted a planetary embryo of $M_{\rm{pl,0}}=10^{-2} M_\oplus$ at different disk times and radial positions. We assumed the embryo is formed through other processes beyond the scope of this work. One possible mechanism for embryo formation is the streaming instability, in which high-density filaments of material collapse into planetary embryos \citep{YoudinGoodman05,Johansen+07,Johansen+15}. Here, we set aside planet migration and did not track the chemical composition of the building blocks. This will be investigated in future works.\par
A planet exerts a gravitational influence in the disk over a distance according to its Hill radius,
\beq
r_{\rm{Hill}} = r_{\rm{pl}}\left(\frac{M_{\rm{pl}}}{M_*}\right)^{1/3},
\eeq
where $r_{\rm pl}$ and $M_{\rm pl}$ are the radial planet location and the planet mass, respectively. We opted to insert Moon-mass embryos into our disk, as they accrete pebbles more effectively. The prescription for the most efficient 2D pebble accretion is given by \citep{Ormel17}
\beq
\label{eqn:pebble_accretion}
\dot{M}_{\rm{p}} = 2  r_{\rm{Hill}}^2 \Omega \rm{St}^{2/3} \Sigma_{\rm{p}}.
\eeq
\subsection{Planet--disk interactions}
\label{sect:planet_disk_interactions}
As the planet grows, it also begins to exert a tidal force on the disk. This torque can clear a gap in the gaseous disk, depending on the planet mass. We used the formulation in \citet{LinPapaloizou86}, where the momentum exchange between the planet and the disk allows the planet to clear a gap in the gas surface density. This effect translates into a change in the gas velocity on on both sides of the planet location. We call this velocity $u_{\tau}$, and it is given by
\beq
u_{\tau} = \frac{f_{\Lambda}\sqrt{GM_\odot r}}{r}\left(\frac{M_{\rm pl}}{M_\odot}\right)^2 \left(\frac{r}{{\rm max}(|r-r_{\rm{pl}}|, H)}\right)^4
,\eeq
where $f_\Lambda$ is a numerical constant.
This velocity either accelerates or diminishes the flow of pebbles, depending on the location in the disk. Pebbles located between the planet and the star are accelerated toward the star, while pebbles beyond the planet are accelerated toward the outer disk. This reversal in the gas velocity leads to a halt in the pebble drift and the formation of a pressure bump outside of the planet location. This behavior prevents pebbles from entering the planet location, halting the planet formation process. Recent studies of planet formation used the pebble isolation mass to calculate the end of growth via pebble accretion \citep{Lambrechts+14, Bitsch+18}. We investigated how a gradual stop of the formation process affects the evolution of pebble dynamics and super-Earth formation in 1D models.
\subsection{Modeling setup}
We evolved our disk for 3 Myr using the finite-volume approach, which allowed us to use a complex and computationally fast model. The model's computation time was determined by the time step of the model. We limited the time step based on the three disk populations. We ensured that the gas, dust, and pebble densities did not change by more than 0.1\% of their previous value. This results in the best stability for disk evolution and mass conservation.
\subsection{Model parameters}
\begin{table}
  \caption{Parameters used throughout this paper.}
  \label{tab:params}
  \centering
    \begin{tabular}{ | l || c | l | }
      \hline
      \multicolumn{3}{|c|}{Disk Parameters} \\
      \hline
      Parameter & Explanation & Value \\
      \hline
      $\gamma$ & power-law exponent & 0.8 \\
      $r_c$ & critical disk radius & 60 AU  \\
      $M_{\rm{disk}}$ & disk mass & $0.1 M_\odot$\\
      $\mu$ & average molecular mass & 2.4 \\
      $\varphi$ & grazing angle & 0.05 \\
      $L_*$ & stellar luminosity & $1 \ L_\odot$  \\
      $M_*$ & stellar mass &  $1 \ M_\odot$ \\
      $Z_0$ & initial dust-to-gas ratio & 0.01  \\
      $\kappa_{\rm{R}}$ & Rosseland mean opacity & 508 \\
      $\alpha$ & alpha viscosity & [$10^{-4} - 10^{-2}$] \\
      \hline
      \multicolumn{3}{|c|}{Planet/Pebble Parameters} \\
      \hline
      $M_{\rm{pl,0}}$ & planetary embryo mass & $10^{-2} \ M_\oplus$ \\
      $t_0$ & initial time & $[5 \times 10^4 - 10^6] \ \rm{years}$ \\
      $\rho_{\rm{s}}$ & grain material density & 1.25 $\rm{[g / cm^{3}]}$ \\
      $a_0$ & initial dust size & $10^{-4} \ \rm{[cm]}$ \\
      \hline
    \end{tabular}
\end{table}

%
%
We defined our disk quantities on two logarithmically spaced grids. The first comprised $N_1=200$ points between $r_{\rm{in}}$ = 0.2 AU and $r_{\rm{out}}$ = 1000 AU. The second grid comprised $N_2=200$ points 0.5 AU from the planet in both directions (toward and away from the star) to accurately capture the gap carving. The code then solves the disk structure, calculates the smallest time step from the previously described quantities, and advances the model in time.
We began with a solar-mass star and a surrounding disk of $0.1 \ M_\odot$. We varied the viscosity in our disk from $\alpha = [10^{-4}-10^{-2}]$ on a logarithmically spaced grid of 7 points. We also investigated the effect that embryo formation times have on the final planet mass, using a logarithmically spaced grid of 20 points from $t_{0} = [5\times10^{4} - 1\times10^{6}]$ years. Our initial disk parameters are shown in Table \ref{tab:params}. We inserted the individual protoplanets at 0.25, 1, 2, 5, and 10 AU.
\section{Results}
\label{sect:results}
We began by investigating the implications of our two different temperature models. We ran a set of six simulations that varied our two temperature prescriptions and three different viscosities without inserting planetary embryos. We first compared the temporal evolution of the disk temperature for the two temperature prescriptions, and then quantified the effect on the pebble surface density. Lastly, we ran an array of planet formation models to visualize the effects of different pebble surface densities on growing planets. For this, we ran 140 models for 7 different viscosities and 20 embryo insertion times for each temperature prescription. We reran these simulations at radial locations of 0.25, 1, 2, 5, and 10 AU.
\subsection{Differing temperature treatments}
\begin{figure}[h]
  \begin{center}
    \resizebox{\columnwidth}{!}{\includegraphics{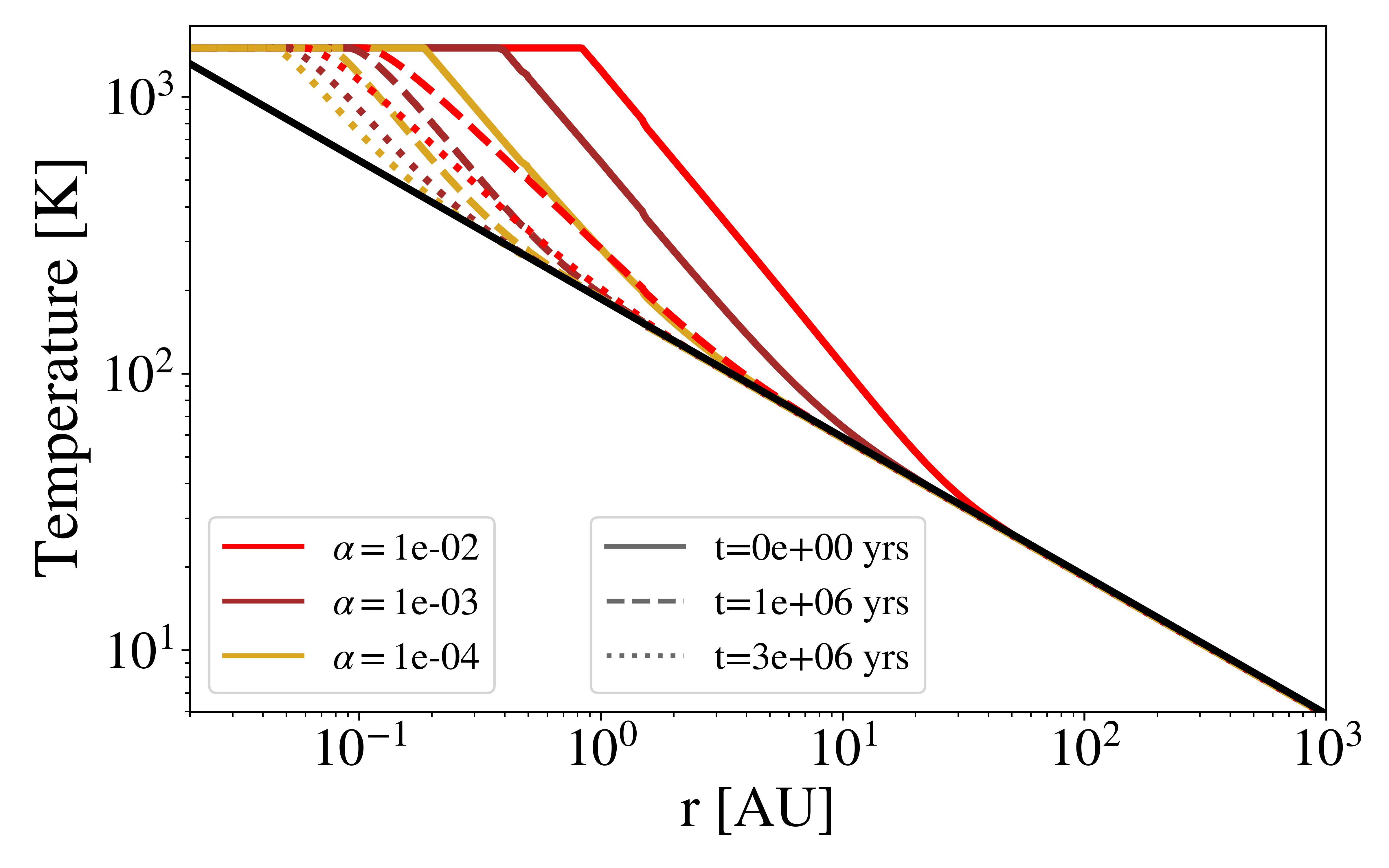}}
    \caption{Temperature structure of the disk as a function of distance from the central star. The solid red, brown, and orange lines correspond to the initial temperature profiles using viscosities of $10^{-2}$, $10^{-3}$, and $10^{-4}$, respectively. The solid black line is a power-law temperature profile that remains static, chosen to match the outer disk temperature of a passively heated disk. The dashed and dotted lines show the evolution of the temperature at 0, $10^6$, and $3\times10^6$ years.}
    \label{fig:temp_comp_powerlaw}
  \end{center}
\end{figure}

We first investigated the differences between temperature prescriptions and their evolution over the disk's lifetime. We began by varying the viscosity for each formulation. Figure \ref{fig:temp_comp_powerlaw} shows the evolving temperature structure for two prescriptions. Larger viscosities have a noticeable effect on the disk temperature in the inner disk. Initially, the value of $T$ at 1 AU varies by a factor of five between viscosities of $10^{-2}$ and $10^{-4}$. Higher temperatures also extend farther into the disk, with the higher alpha value deviating from the power-law case by up to 30 AU. After $3\times10^6$ years, the temperature is reduced for all viscosities, and viscous heating only remains important within 1 AU of the star. This change in temperature can have implications for temperature-dependent quantities, such as pebble drift.
\subsection{Effect on pebble surface density}
\label{sect:drift}
\begin{figure}[h]
  \begin{center}
    \resizebox{\columnwidth}{!}{\includegraphics{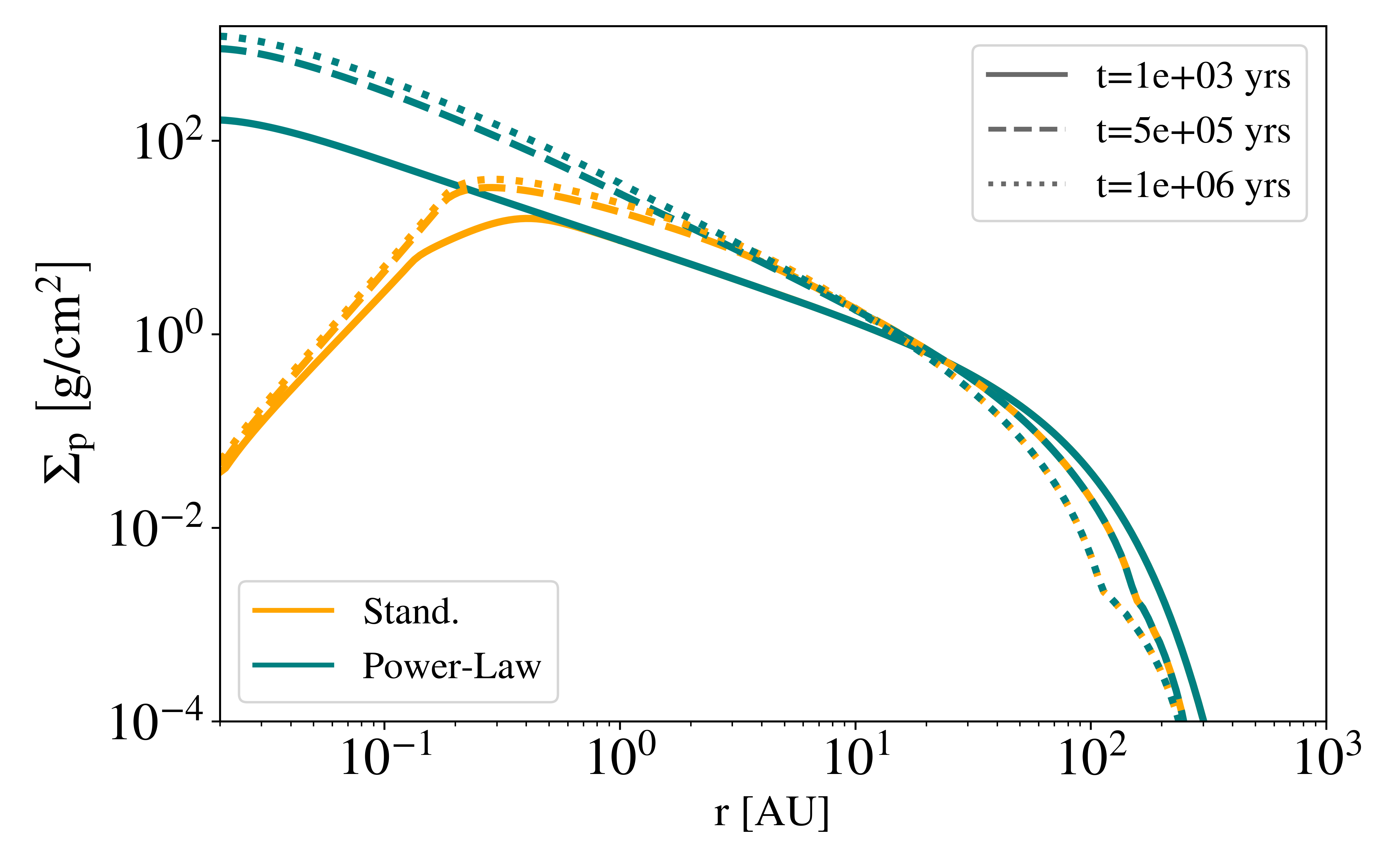}}
    \caption{Temporal evolution of the pebble surface density as a function of radius. The solid orange and teal lines correspond to the pebble surface density for our standard and power-law cases, respectively. The orange lines show the dust-sublimation front beginning at around 0.5 AU. The solid, dashed, and dotted-dashed lines correspond to disk times of $10^3$ years, $5 \times 10^5$ years, and 1 Myr, respectively.}
    \label{fig:pebble_comp_temp}
  \end{center}
\end{figure}

One of the evolving quantities affected by temperature is the pebble surface density. Figure \ref{fig:pebble_comp_temp} shows the evolution of pebbles as a function of radius up to $10^6$ years in the absence of protoplanets. We chose to show the intermediate case of an alpha value equal to $10^{-3}$. At $t=1\times10^3$ years, the pebble surface densities for the standard and power-law cases are similar throughout the disk, except for the location of the dust sublimation front at 0.5 AU, where the standard case shows a reduced pebble surface density. The reduction in this part of the disk is due to the removal of dust, which in turn is due to the thermostat effect at $T=1500 \ {\rm K}$; this results in decreased formation of pebbles due to the lack of dust available. Initially, the power-law model shows an increased surface density at the inner edge of the disk, differing by three orders of magnitude from the standard model. As the disk evolves, the pebble surface density decreases in all cases, as inward drift removes pebbles through the inner boundary. At $t=5\times10^5$ years, the pebble surface density for the standard treatment and the power-law case show a reduction of pebbles in the outer disk and an enhancement of pebbles in the inner disk. We note that the surface densities at 1 AU differ, with the power law having a smaller enhancement than our standard case. By $t=10^6$ years, the outer pebble density has migrated inward, leading to a pebble-rich disk inside 10 AU. We note that the largest difference in the pebble surface densities occurs where dust sublimation becomes significant. Star-ward of this location, the difference in the pebble surface density is up to four orders of magnitude.
%
\begin{figure}[t]
  \begin{center}
    \resizebox{\columnwidth}{!}{
      \includegraphics[]{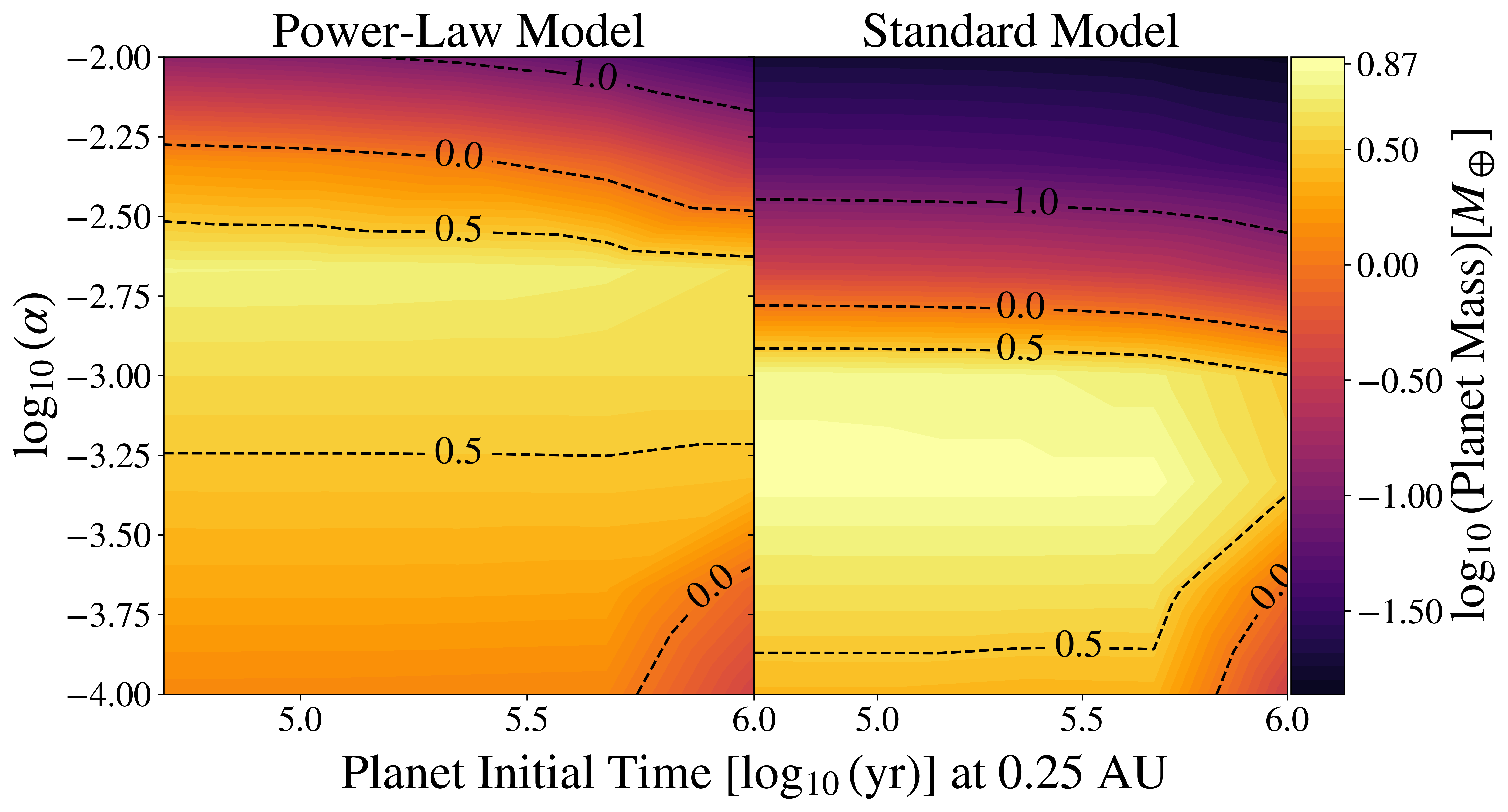}
    }
  \end{center}
  \caption{Planet formation grid at 0.25 AU. We show the final planet masses after $3\times10^6$ years of evolution as a function of alpha viscosity and the time a planetary embryo is inserted in the disk. Left panel: Power-law case with static irradiative heating. Right panel: Standard model with a dust-sublimation front that causes a temperature cap. The contours in black show the final planet masses differing by orders of 0.5 dex.}
  \label{fig:comp_025au}
\end{figure}
\begin{figure}[t]
  \begin{center}
    \resizebox{\columnwidth}{!}{
      \includegraphics[]{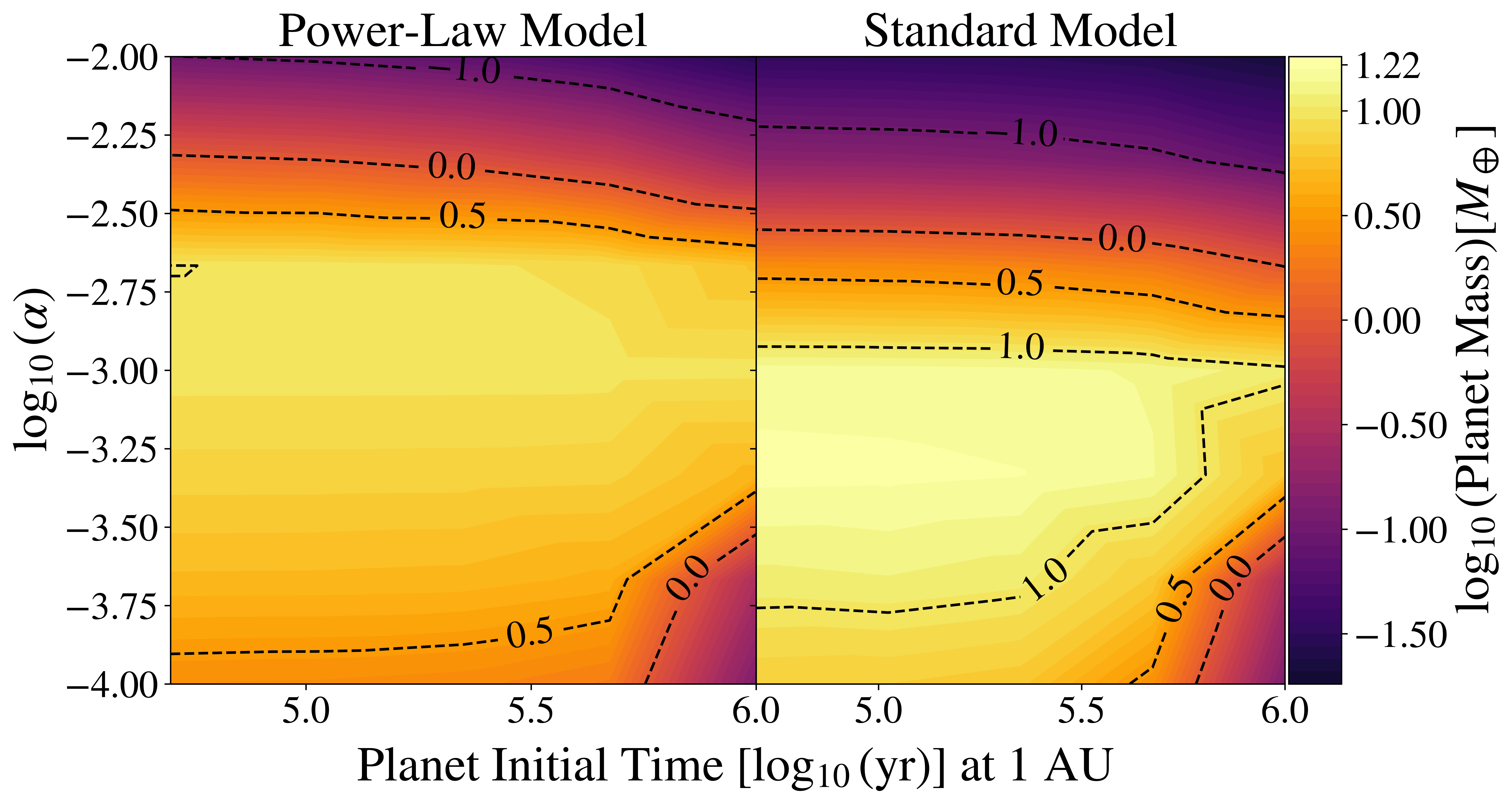}
    }
  \end{center}
  \caption{Same as \fig{fig:comp_025au} but at 1 AU in the disk.}
  \label{fig:comp_1au}
\end{figure}
\begin{figure}[]
  \begin{center}
    \resizebox{\columnwidth}{!}{
      \includegraphics[]{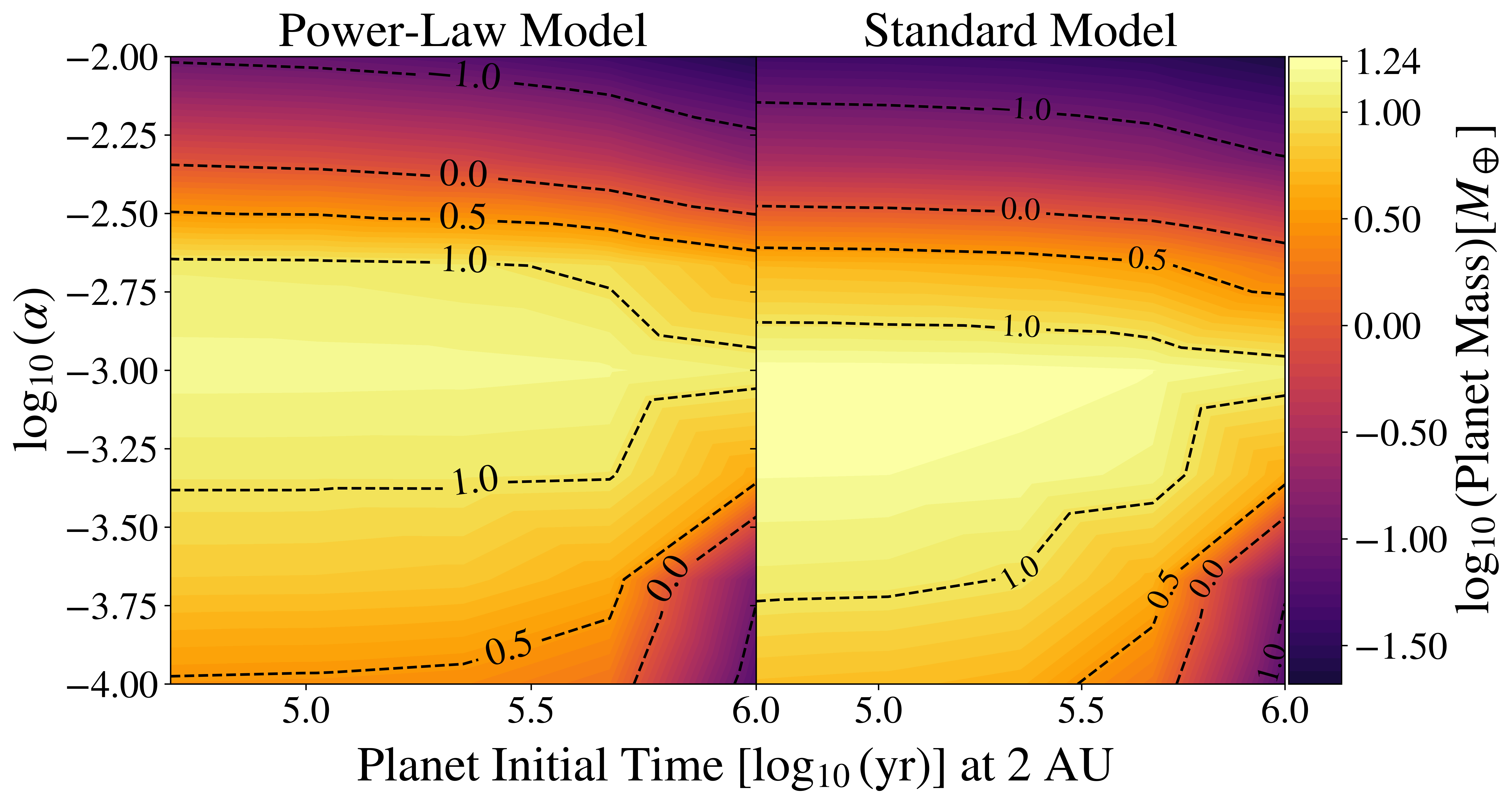}
    }
  \end{center}
  \caption{Same as \fig{fig:comp_1au} but at 2 AU in the disk.}
  \label{fig:comp_2au}
\end{figure}
\begin{figure}[]
  \begin{center}
    \resizebox{\columnwidth}{!}{
      \includegraphics{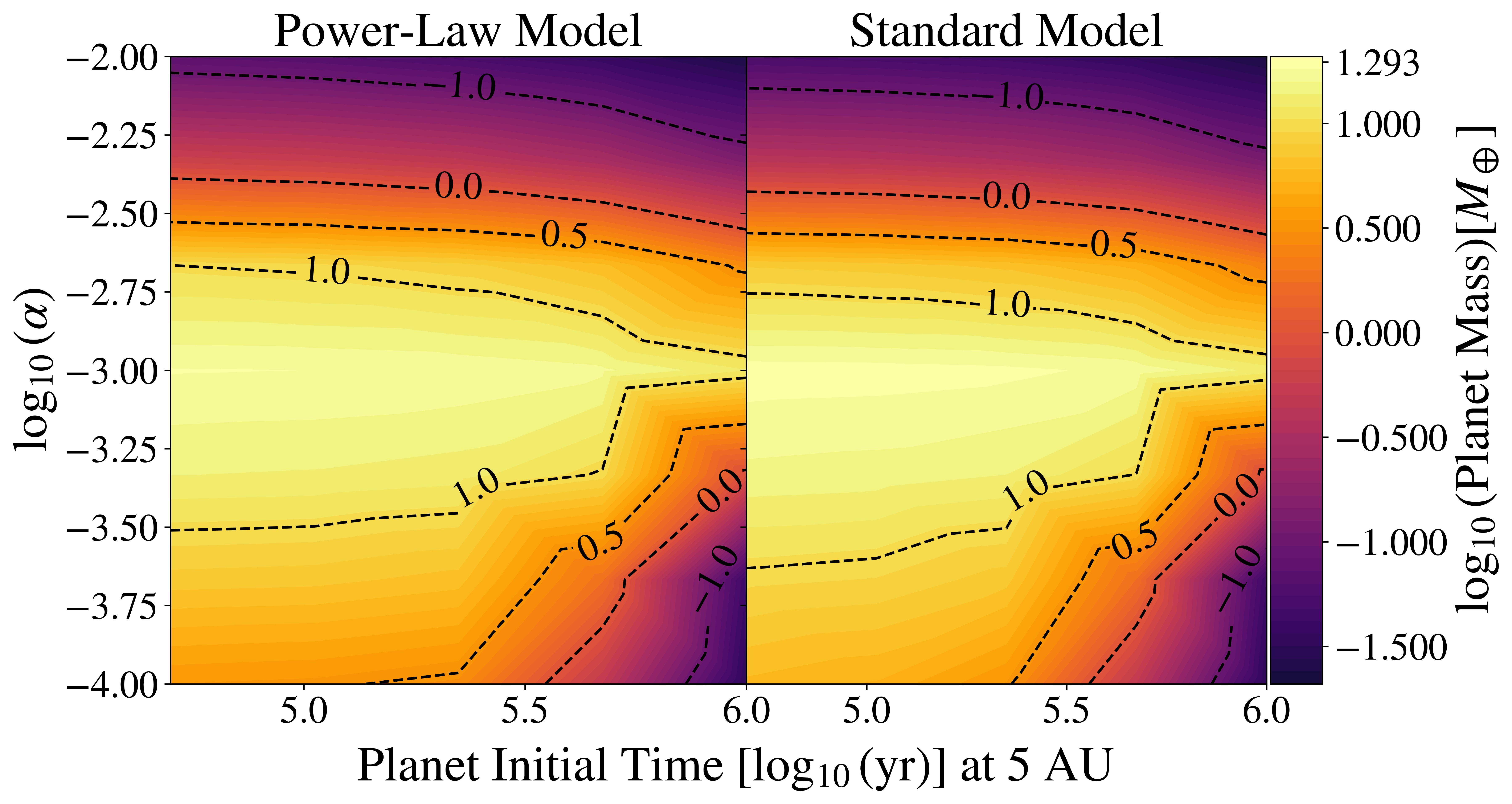}
    }
  \end{center}
  \caption{Same as \fig{fig:comp_1au} but at 5 AU in the disk.}
  \label{fig:comp_5au}
\end{figure}
\begin{figure}[]
  \begin{center}
    \resizebox{\columnwidth}{!}{
      \includegraphics{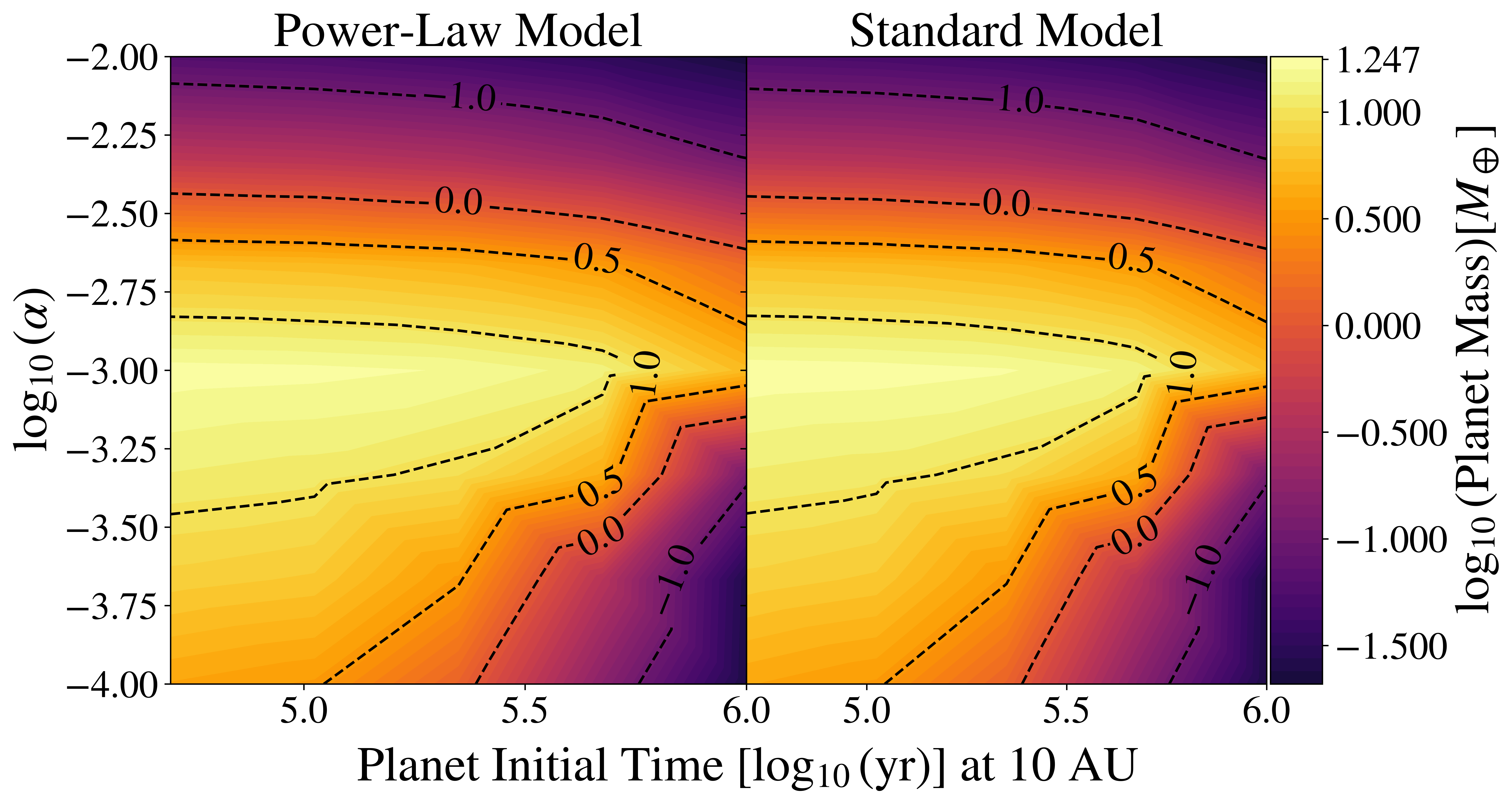}
      }
  \end{center}
  \caption{Same as \fig{fig:comp_1au} but at 10 AU in the disk.}
  \label{fig:comp_10au}
\end{figure}
\subsection{Effect on planet formation}
In the previous section, we explored the effect of the different temperature prescriptions on the evolution of the pebble surface density in the absence of planets. We found that with a non-evolving temperature structure, there was more pebble availability inside the dust-sublimation front. Since pebbles form the basis of planet growth in our model, we expected the effect on planet populations to be nonzero. We tested this by placing planetary embryos at 0.25, 1, 2, 5, and 10 AU. We varied the initial time at which the planetary embryo was inserted. The embryos then grew through pebble accretion until we stopped the disk evolution at 3 Myr.\par
Figure \ref{fig:comp_025au} shows the final masses for planets formed at 0.25 AU. The location of these planets is within the dust-sublimation front, where the availability of dust and pebbles was reduced. The noticeable difference between the models is that with the standard model, planets do not grow larger than 1 Earth mass at alpha values larger than $10^{-3}$, whereas the power-law model shows a distribution of planets with masses larger than that of Earth across all alpha values. At this location, the largest planets formed are 7 Earth masses for our standard case and 6 Earth masses for the power-law case. Both panels also show preferential growth for alpha values of $10^{-3}$ in the power-law case, whereas the standard case shows a preference for alpha values below $10^{-3}.$\par
Figure \ref{fig:comp_1au} shows the final masses for the different planets formed at 1 AU. At this location, the standard model shows more massive planets than the power-law case. The most massive planet formed is 16 Earth masses. We find a correlation between earlier growth times and more massive planets, with the majority of planets lying between 1 and 16 Earth masses. High alpha values of $\approx \alpha=10^{-2}$ do not allow for the formation of planets larger than 1 Earth mass. This is due to the gaseous disk spreading for larger alphas, dragging pebbles along, and reducing the amount of planet-forming material in the disk. This is also coupled with the fact that higher alphas lead to lower particle sizes and Stokes numbers due to fragmentation. Particles with lower Stokes numbers lead to inefficient pebble accretion. The figure shows a preference for planet formation at intermediate alpha values. This is explained in detail in Sect. \ref{subsect:alpha_sweet_spot}. The fixed power law shows a similar distribution of planet masses as a function of viscosity and embryo insertion time. However, they are generally 40-60 \% lower in mass compared to the standard case.
If the embryo is placed at 2 AU (\fig{fig:comp_2au}), the largest planets formed for our standard temperature case are on the order of 17 Earth masses. This is a slight increase in mass from planets formed at 1 AU. In the power-law case, the largest planet is on the order of 15 Earth masses. However, we note that the power-law case can now form planets larger than 10 Earth masses at this orbital separation.\par
Next, we placed a planetary embryo at 5 AU (\fig{fig:comp_5au}). The standard temperature model continues to show a varied distribution of planet masses larger than Earth, with the most massive being 20 Earth masses. The static power-law model shows a similar distribution of planets, with the largest being 18 Earth masses. Therefore, larger orbital separations result in a marginal increase in planetary masses. This is to be expected, as the Hill radius for a Moon-mass embryo grows more the farther out it is placed in the disk.\par
Lastly, we placed a planetary embryo at 10 AU (\fig{fig:comp_10au}). In all models, the largest planets are $\approx$ 17 Earth masses, falling within the mini-Neptune regime. The mass difference in the largest planets at this location is less than 0.5 Earth masses. At this orbital separation, we find that both cases are similar in their planet production.\par
Thus, we find that a self-consistent evolving temperature structure results in differences in the pebble surface evolution, which translates into a difference in the final planet mass. Dynamic temperatures at 1 AU lead to an increase of 60\% in the final planet mass for the largest planet. However, as we move radially outward and the dominant temperature becomes the irradiation from the star, the mass increase diminishes to only 3\% at 10 AU. When a planet is placed within the dust-sublimation line, the largest planets are of super-Earth sizes, and the power law shows a larger distribution of super-Earth planets. This is to be expected since the power-law prescription has a greater availability of pebbles for planet formation.\par
\section{Discussion}
\label{sect:discussion}
\subsection{The $\alpha$ "sweet spot"}
\label{subsect:alpha_sweet_spot}
\begin{figure}[h]
  \begin{center}
    \resizebox{\columnwidth}{!}{\includegraphics{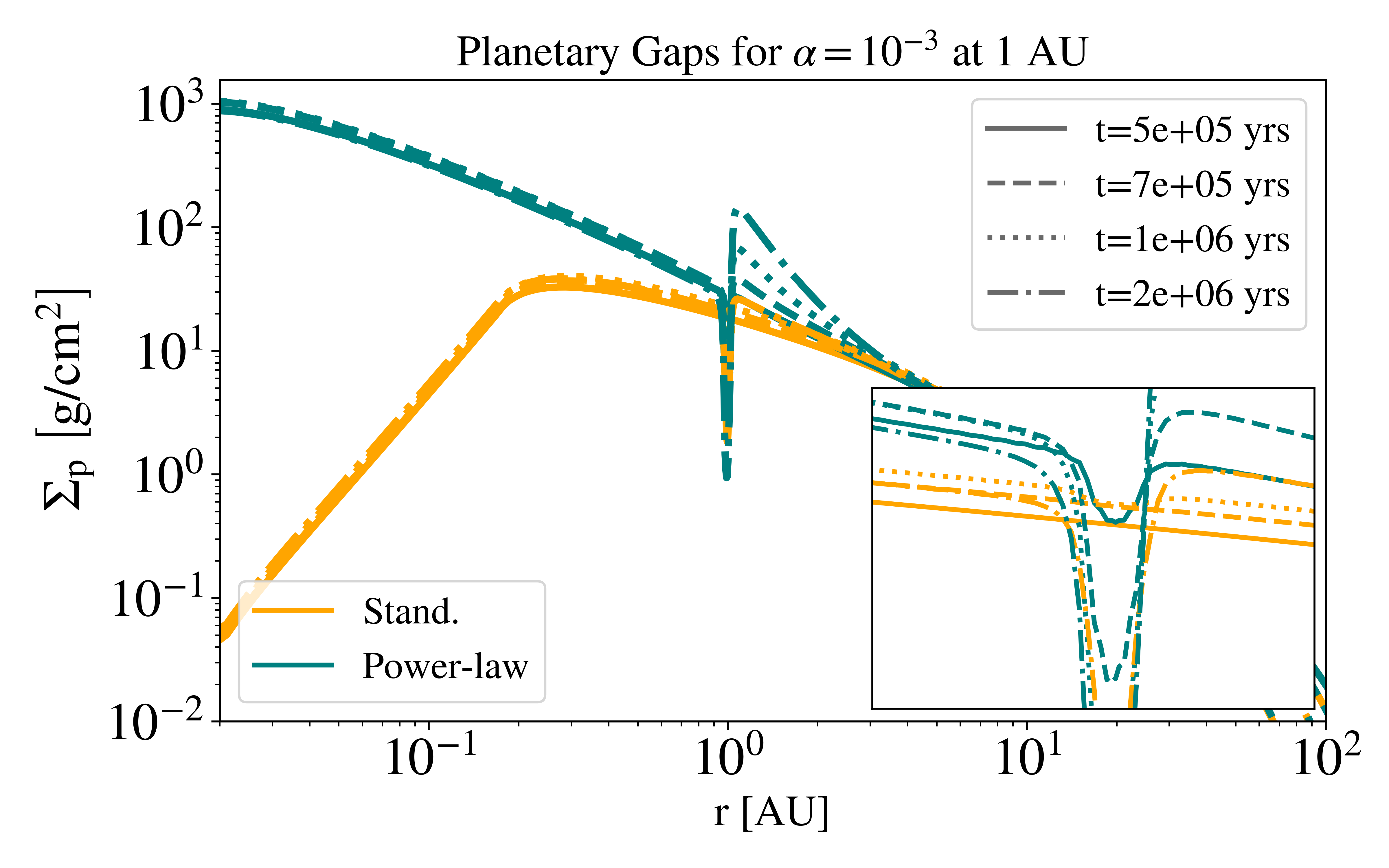}}
    \caption{Gap formation in the pebble surface density by planets in a disk at 1 AU with an alpha value of $10^{-3}$ and an embryo insertion time of $5\times10^4$ years. The orange and teal lines correspond to the standard and power-law temperature structures. The insert shows the growth of the gaps in the pebble surface density at $5\times10^5$, $7\times10^5$, $10^6$, and $2\times10^6$ years.}
    \label{fig:pebble_gap_comp}
  \end{center}
\end{figure}

\begin{figure}[h]
  \begin{center}
    \resizebox{0.75\columnwidth}{!}{\includegraphics{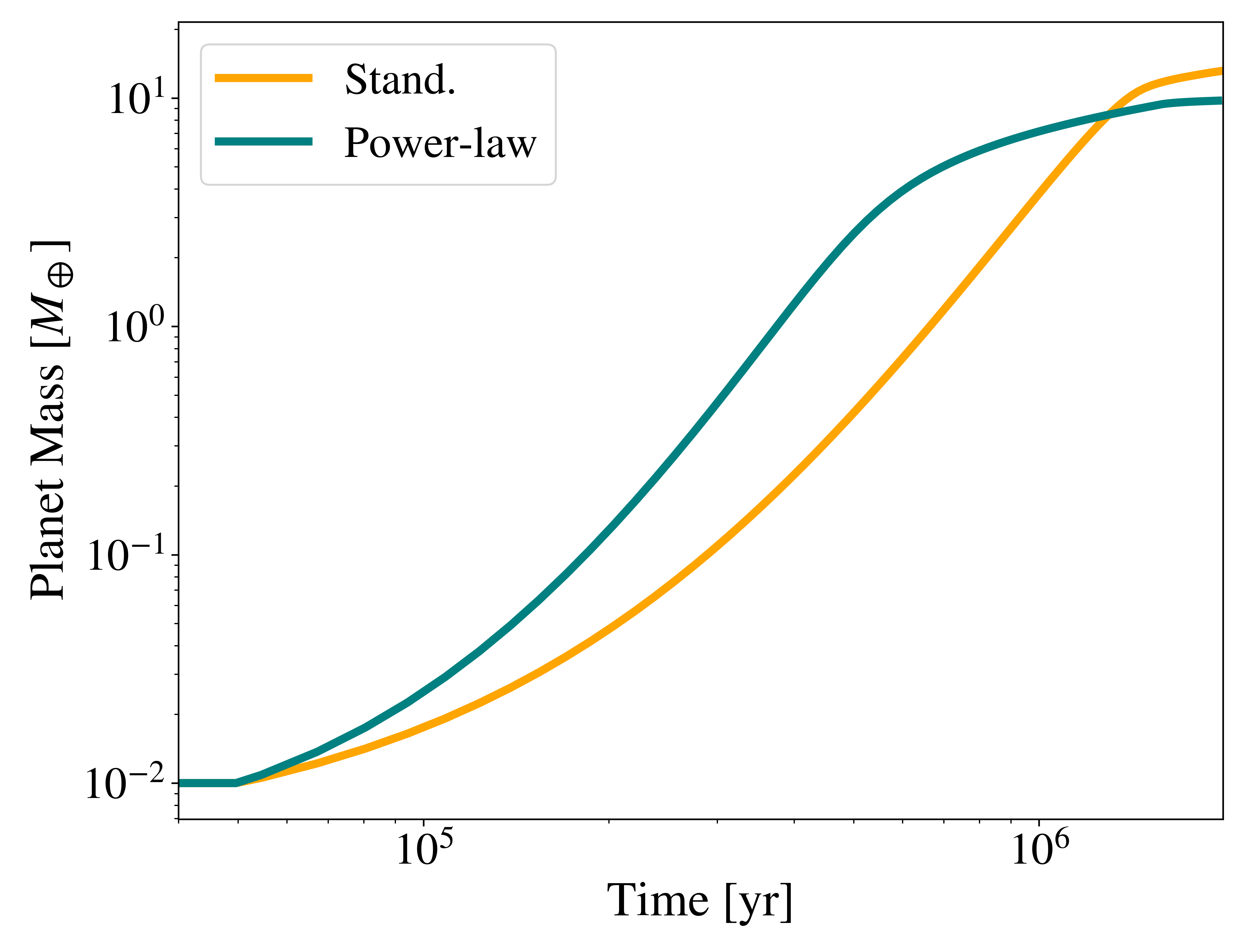}}
    \caption{Growth of a Moon-mass embryo for $\alpha=10^{-3}$ from the insertion time until 2 Myr, when the planet growth plateaus. The orange and teal lines show the evolution of the planet mass for the standard and power-law cases.}
    \label{fig:planet_evolution}
  \end{center}
\end{figure}

At all radial locations in the disk, we find an optimal growth mode for planets when $\alpha$ is equal to $10^{-3}$. This "sweet spot" can be explained by the gap carving in our disk. At one extreme, a low alpha leads to smaller planets carving a significant gap, stopping the flow of pebbles and stunting the planet's growth. At the other extreme, a high alpha impedes a planet from forming a significant gap, and material streams past the planet too quickly, without being accreted. This results in the reduced planet masses at the boundaries of our $\alpha$ parameter exploration. The effect of an intermediate alpha is most noticeable when we compare the two different temperature structures (Fig. \ref{fig:pebble_gap_comp}). In Sect. \ref{sect:drift}, we established that a passively heated disk results in an abundance of pebbles due to lack of sublimation. This enhances pebbles in the inner disk and allows a planet to grow large enough to form a gap by $t = 7 \times 10^5$ years. The insert shows the gap formation for our two temperature prescriptions. The power-law temperature shows a noticeable gap at all times, while the standard case lacks any deviation, even at 1 Myr. After 2 Myr, the planet in the standard case has grown large enough and carves a significant gap in the pebble surface density. Therefore, a dynamic temperature prescription suppresses gap formation in a 1D context. This lack of a gap allows material to enter the planet's Hill radius, continuing the growth to higher masses than those at higher or lower alpha values. This behavior causes the "sweet spot" for planet growth at intermediate alpha values. Figure \ref{fig:planet_evolution} shows that the power-law case grows to its final planet mass more quickly than the standard prescription. This early growth is due to a larger pebble reservoir (Fig. \ref{fig:pebble_comp_temp}). This leads to gap carving at an earlier time, which stunts the growth of the planet.
\subsection{Gas accretion}
Our model produces planets within Earth to mini-Neptune masses. We did not model runaway gas accretion and allowed our planets to grow to masses of $\approx10 \ M_{\oplus}$ solely through the accretion of pebbles. In fact, it has been shown that planets already begin accreting a gaseous envelope in early stages \citep{Ikoma+00}. At 5 Earth masses, the accretion of a gaseous envelope will be nonzero. In our model, we find that inside 10 AU, planets grow to masses high enough to be the seeds for gas giants. Therefore, the effects of a gaseous envelope will have to be included to model the complete planetary demographics. We have left this for a future study.
\subsection{Embryo masses}
Recent studies show that the radial location dictates an embryo's initial mass in streaming instability simulations \citep{Liu+20}. As our focus here was the effect that temperature structures have on pebble accretion and planet formation, we chose to keep the initial embryo mass constant, in order to isolate the influence of temperature fluctuations during this study.\par
\subsection{Gap shape}
\begin{figure}[t]
  \begin{center}
    \resizebox{\columnwidth}{!}{
      \includegraphics{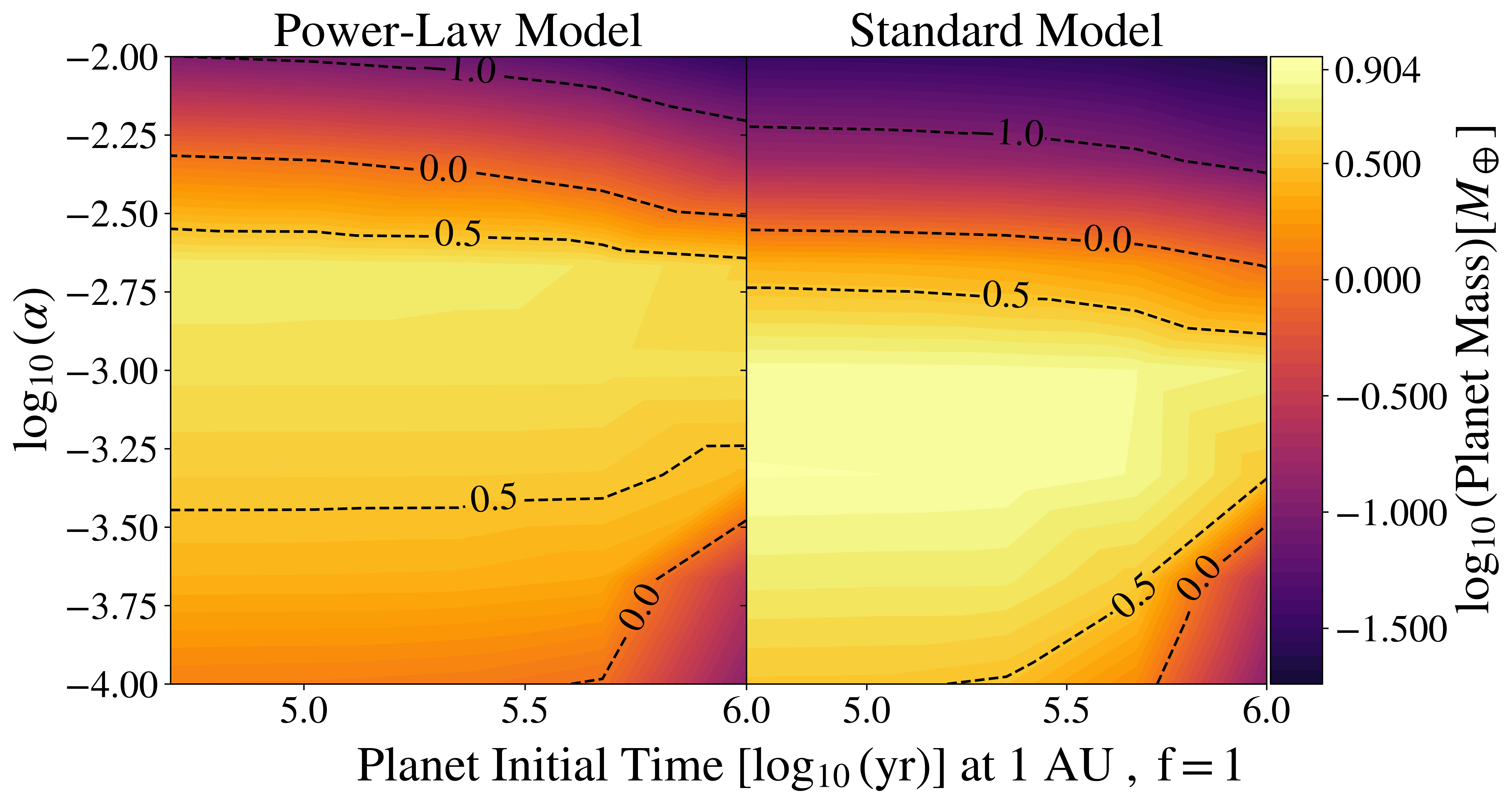}
      }
  \end{center}
  \caption{Same as \fig{fig:comp_1au} but with the torque factor, $f_{\Lambda}$, changed from 0.23 to 1.}
  \label{fig:f1_comp_1au}
\end{figure}

We used the impulse approximation for a planetary torque to form a gap in the gaseous disk by converting it to a gas velocity \citep{LinPapaloizou79}. This caused a change in the pressure gradient, allowing for pebbles to be slowed and subsequently stopped due to the gap formed by a growing planet. Simulations in 2D and 3D show that gaps formed by 1D approximations overestimate the depth of the gap in the disk. The difference can be as large as two orders of magnitude when compared to similar 2D gap-opening models \citep{HallamPaardekooper17}. To account for this, the impulse approximation used contains a factor $f_{\Lambda}$ that is usually taken to be 0.23. We investigated the effect that this factor has on a resulting planetary population. We reran the models at 1 AU using $f_\Lambda=1$ (Fig. \ref{fig:f1_comp_1au}) and compared this to Fig. \ref{fig:comp_1au}. We find that across all parameters, the final mass of planets formed decreases. The largest planets grown with $f_\Lambda=0.23$ are 17 Earth masses, and 8 Earth masses with $f_\Lambda=1$. Modifying this factor shifts the planet population down by a factor of two, meaning it is possible to form more super-Earths and fewer mini-Neptunes. Changing the $f_\Lambda$ by a factor of five causes the planet to remove material close to its Hill radius, effectively halting the planet-forming process at an earlier stage.
\subsection{Migration}
Gap formation due to a planet will also lead to planet migration, depending on the planet mass. We omitted planet migration and are aware of the implications. Migrating planets can relocate inward to areas with larger mass budgets and grow into the runaway gas accretion regime. \citet{Khorshid+21} modeled gas giant evolution after the pebble isolation mass is reached and constrained their migration history. They find that, based on a planet's metallicity and C/O ratio, they can predict the initial planet location. These results mirror the importance of migration in another study that finds that the composition of evaporating pebbles can affect the total C/O ratio in gas giant atmospheres \citep{SchneiderBitsch21}. This migration process should also affect the pebble composition in intermediate-mass planets, as different radial locations will result in varied bulk compositions. The effect of migration is also left for a future study.\par
\subsection{Comparison to other works}
\label{sect:comparison}
Recent works have also examined the role of pebble accretion and the formation of intermediate-mass planets in the super-Earth and mini-Neptune range \citep{Venturini+20,SavvidouBitsch21}. \citet{Venturini+20} used a viscous disk, while including photoevaporation in the inner parts of the disk. They also included planet migration and gas accretion, and ignored 1D gap prescriptions. Intriguingly, in the low-alpha case for our model ($\alpha=10^{-4}$), we see results similar to theirs for in situ planet formation at 1 AU. They report the largest final mass to be 2.6$M_\oplus$, and we find a mass of 7$M_\oplus$; there is better agreement when we take the torque factor, $f_\Lambda$, to be unity, with our highest planet mass being 3$M_\oplus$, as seen in Fig. \ref{fig:f1_comp_1au}. An increase in $f_\Lambda$ leads to a stronger planetary torque arresting the planet growth at earlier times. The difference in the resulting planet masses shows that our 1D gap prescription overestimates the planet growth by a factor of two when compared to 2D simulations.\par
\citet{SavvidouBitsch21} explored the formation of super-Earths using 2D simulations in a limited disk region between 0.1 and 4 AU, without gaps, while ignoring planet migration and gas accretion. These models were evolved for a shorter amount of time, until the disk reached thermal equilibrium. Our model has the potential to run for timescales of up to millions of years, allowing us to model the disk evolution over a longer time. \citet{SavvidouBitsch21} took a self-consistent approach to opacities, varying the opacities of various dust species as a function of temperature. They find that the optimal alpha value for the formation of planets large enough to reach the pebble isolation mass is $\alpha=10^{-3}$. Interestingly, with different assumptions and a different modeling setup, we find the same optimal value for alpha. We find that the formulation in our model is sufficient.
%
%
\section{Conclusion}
\label{sect:conclusions}
In this work, we developed a semi-analytic model of a protoplanetary disk. We aimed to study the effect that different temperature prescriptions have on the evolution of gas and solids in the disk over time. Knowing the effect of the temperature, we set out to test the effect on a population of pebbles. 
We find that a power-law prescription initially corresponds to an abundance of pebbles when compared to a dynamic temperature, due to the lack of dust sublimation (Fig. \ref{fig:pebble_comp_temp}). We find that a time-dependent temperature treatment changes the availability and dynamics of pebbles in a significant way. 
We ran an array of models, varying the disk viscosity, initial embryo formation time, and planetary location. We find that there is a preferred mode for forming larger planets around $\alpha=10^{-3}$, resulting in planets on the order of $\approx 10 M_\oplus$.
\par
To conclude, we find that including viscous heating in an evolving disk leads to changes in the pebble surface density and gap-formation mechanisms. The changes in these two quantities result in planets with higher masses where viscous heating is the primary heating mechanism. Lastly, we find that the inclusion of a dust-sublimation front is beneficial for the growth of super-Earths at a short orbital separation. This could be a possible explanation for the large occurrence of super-Earths at close-in orbits seen in \textit{Kepler} data.
%
%
\begin{acknowledgements}
We thank the anonymous referee for their comments, which greatly improved the manuscript. We acknowledge funding from the European Union H2020-MSCA-ITN-2019 under Grant no. 860470 (CHAMELEON). U.G.J. also acknowledges funding from the Novo Nordisk Foundation Interdisciplinary Synergy Programme grant no. NNF19OC0057374.
\end{acknowledgements}
\bibliography{aa47347-23corr}{}
\end{document}